\documentclass[a4paper,11pt]{article}

\pdfoutput=1

\usepackage{jcappub}
\usepackage{amssymb,amsmath,enumerate,caption}
\usepackage{subfig}
\usepackage{graphicx}

\usepackage[T1]{fontenc}

\newcommand{\be}{\begin{equation}}
\newcommand{\bea}{\begin{eqnarray}}
\newcommand{\eea}{\end{eqnarray}}

\newcommand{\beq}{\begin{equation}}
\newcommand{\eeq}{\end{equation}}

\newcommand{\comment}[1]{}

\newcommand{\ee}{\mathrm e}
\newcommand{\mP}{m_{\mathrm P}}
\newcommand{\rd}{\mathrm d}

\newcommand{\phic}{\phi_{\mathrm c}}
\newcommand{\phii}{\phi_{\mathrm i}}
\newcommand{\phif}{\phi_{\mathrm f}}
\newcommand{\Ne}{N_{\mathrm e}}
\newcommand{\Cphi}{C_\phi}
\newcommand{\CphiLM}{C_\phi^{\rm LM}}
\newcommand{\CphiPole}{C_\phi^{\rm pole}}

\newcommand{\Gamchi}{\Gamma_\chi}
\newcommand{\Nchi}{N_\chi}
\newcommand{\mchi}{m_\chi}

\newcommand{\rhoR}{\rho_{\mathrm R}}

\newcommand{\ns}{n_{\rm s}}
\newcommand{\Ups}{\Upsilon}

\newcommand{\ceff}{c_{\rm eff}}

\newcommand{\PR}{P_{\mathcal R}}

\newcommand{\curlO}{\mathcal{O}}

\def \Upsilon {\varUpsilon}

\makeatletter
\newcommand\CMmathsymbol[2][\mathord]{%
  #1{\CM@mathsymbol@i{#2}}%
}

\def\CM@mathsymbol@i#1{\mathchoice
  {\CM@mathsymbol@ii{#1}\tf@size}%
  {\CM@mathsymbol@ii{#1}\tf@size}%
  {\CM@mathsymbol@ii{#1}\sf@size}%
  {\CM@mathsymbol@ii{#1}\ssf@size}%
}

\def\CM@mathsymbol@ii#1#2{%
  \mbox{\fontsize{#2}{#2}\usefont{OT1}{cmr}{m}{n}\symbol{#1}}%
}

\renewcommand{\Upsilon}{\CMmathsymbol[\mathalpha]{07}}
\makeatother
\title{Exploring the Parameter Space of Warm-Inflation Models}
\date{\today}

\author[a]{Mar Bastero-Gil} \emailAdd{mbg@ugr.es} \affiliation[a]{Departamento de
  F\'{\i}sica Te\'orica y del Cosmos, Universidad de Granada, Granada-18071,
  Spain}

\author[b]{Arjun Berera} \emailAdd{ab@ph.ed.ac.uk} \affiliation[b]{SUPA, 
School of Physics
and Astronomy, University of Edinburgh, Edinburgh, EH9 3JZ, United Kingdom}

\author[b]{Nico Kronberg} \emailAdd{nico.kronberg@ed.ac.uk}

\abstract{

Warm inflation includes inflaton interactions with other
fields throughout the inflationary epoch instead of confining such
interactions to a distinct reheating era. Previous investigations have shown that, when certain constraints on the dynamics of these
interactions and the resultant radiation bath are satisfied, a
low-momentum-dominated dissipation coefficient $\propto T^3/\mchi^2$
can sustain an era of inflation compatible with CMB observations. In
this work, we extend these analyses by including the pole-dominated
dissipation term $\propto \sqrt{\mchi T} \exp(-\mchi/T)$. We find
that, with this enhanced dissipation, certain models, notably the
quadratic hilltop potential, perform significantly better.
Specifically, we can achieve 50 $\ee$-folds of inflation and a spectral index compatible with Planck data while requiring fewer
mediator field ($\curlO(10^4)$ for the quadratic hilltop potential)
and smaller coupling constants, opening up interesting model-building possibilities. We also highlight the significance of
the specific parametric dependence of the dissipative coefficient
which could prove useful in even greater reduction in field content.

}
\arxivnumber{}

\begin{document}

\maketitle

\section{Introduction}

Recent CMB data has made it evident that dissipation and particle
production may have a role to play in the inflationary phase.
The lack of detection of a tensor mode has meant that now the
tensor-to-scalar ratio is low enough to rule out two of the most
compelling cold-inflation models, the chaotic $\phi^2$ and
$\phi^4$ inflation 
models \cite{Planck13Inflation,Planck15Params,Martin:2013nzq,Tsujikawa:2013ila,Enqvist:2013eua,Amoros:2015eva}.  
Of course, various fix-ups to these
models are possible that have some limited
success \cite{Barvinsky:2008ia,Hertzberg:2010dc,Tsujikawa:2013ila,Enqvist:2013eua, Kallosh:2013hoa}, but the basic argument that
has kept these models in favor, that of simplicity, has now been
lost.  The warm inflation realization of both these models allows
the tensor-to-scalar ratio to go down to levels constrained by CMB data,
although only the $\phi^4$ model is also consistent with the
bounds on tilt \cite{Bartrum13,BasteroGil:2014oga,BGBMR14}.

The warm-inflation realization of
these models relies on the coupling of the inflaton to other fields,
and the subsequent non-equilibrium dissipative dynamics that
develop from these interactions, leading to particle production
during inflation and to thermal seeds of density perturbations.
Of course, the inflaton field is always coupled 
to other fields \cite{Berera:1995wh,Berera:1995ie,Berera:1999ws,Abbott:1982hn,Dolgov:1982th,Albrecht:1982mp,Kofman:1994rk,Cook:2015vqa}.
Even in cold inflation, this is necessary for the reheating phase
that is meant to follow inflation.  Nevertheless, warm inflation
is rather more technically complicated
than generic cold inflation models. This is because in warm inflation
the quantum-field-theory dynamics of particle production must coincide with inflation, which imposes various demands on the underlying 
dynamics \cite{Berera:1998gx,BasteroGil:2010pb,BasteroGil:2012cm}.
However, from a theoretical
perspective, the couplings in these warm-inflaton realizations are
generic and, aside from requiring global SUSY to cancel radiative
corrections, no other new physics is required beyond what has already
been tested and verified in collider experiments.  Cold
inflation realizations of the chaotic models now consistent
with CMB data require more novel new physics, adjusting the nature of gravity
such as in the Higgs inflation 
model \cite{Barvinsky:2008ia,Bezrukov:2009db} and other
models involving non-minimal 
coupling \cite{Hertzberg:2010dc,Germani:2010gm}.
Other cold inflation models which have become popular since
recent CMB data, such as the Starobinsky model \cite{Starobinsky:1980te}, 
also require rather
novel speculations about the nature of gravity at high energies.
The difficulty with such models is, they are quite contrived, thus
have limited scope for predictiveness. On the other hand, warm-inflation
models don't make radical demands on new physics but are quite
complex.  What is clear though, that neither of the two options
between warm- and cold-inflation dynamics is more compelling at this stage.
And whereas cold-inflation models have been exhaustively studied over
three decades by many authors, there has been relatively little study of warm inflation.

In this paper, we will examine a variety of warm-inflation models
and test their agreement with observation.  There have been various
studies of warm-inflation models and their observational predictions.
One of the features of warm-inflation models constructed from
first-principles quantum field theory has been that after all the constraints
are applied, the models usually work at very high field content.
The reason for this is not, as one might naively assume, that
more fields implies more channels for dissipation, thus more
radiation.  Instead, large numbers of fields arise from
requiring the first-principles model to satisfy both the usual observational constraints and consistency constraints from the field theory.  The success of
warm inflation models with the observational data is a significant
result, but now we would like to further understand the underlying
dynamics and the constraints involved and see if better parametric regimes
can be obtained, in particular with lower number of fields required,
to realize observationally consistent warm inflation.

This paper will therefore attempt a more in-depth analysis of the parameter
space in a variety of warm-inflation models based on monomial, hybrid, and hilltop potentials with different powers of the inflaton field. For this, we will
use numerical algorithms that allow exploration of
the parameter space in search of regimes consistent with observation
and the theoretical constraints.

We will scan over the 6-dimensional parameter space that sets the coupling constants and field content of these models as well as the initial conditions for radiation-density and inflaton evolution. For each randomly generated combination of these parameters, we check the conditions necessary for slow-roll and for warm inflation; we then integrate the coupled set of evolution equations for the inflaton, the radiation density, and the scale factor until any of these conditions break down or until the radiation density comes to dominate over the potential energy of the inflaton.

In section~\ref{sec:WIintro} we briefly introduce the dynamics of warm inflation with a general dissipation coefficient; a more thorough review can be found in ref.~\cite{BasteroGil:2012cm}. Section~\ref{sec:Observables} presents the spectral index and the tensor-to-scalar ratio for this class of models; our models' predictions for these observables will be fundamental to our comparison to Planck results in later sections. Section~\ref{sec:mchiTconst} argues that for certain powers of the inflaton field in the potential, the ratio between the field value and the temperature is constant during slow-roll; these models then form part of the numerical investigations we describe in section~\ref{sec:Numerics}. We present our results and conclusions in sections~\ref{sec:Results} and~\ref{sec:Conclusions}.

\section{Warm inflation with general dissipation coefficient}\label{sec:WIintro}

During warm inflation, a small part of the inflaton's energy is dissipated into other fields; in a supersymmetric model, this can be accomplished by the superpotential \cite{BasteroGil09,BasteroGil:2012cm}
\be
W = \frac{g}{2} \Phi X^2 + \frac{h_i}{2} X Y_i^2 + f(\Phi)\,. \label{eq:superpot}
\eeq
In this model, the inflaton $\phi$ is given by the scalar component of the chiral superfield $\Phi$ with expectation value $\langle |\Phi| \rangle = \phi/\sqrt2$; dissipation is mediated by the coupling $g$ to the bosonic and fermionic components of the chiral superfield $X$, labeled $\chi$ and $\psi_\chi$, respectively. Via the couplings $h_i$, these heavy $\chi$ fields decay into the bosonic $y_i$ and fermionic $\psi_{y_i}$ components of the chiral superfield $Y_i$. We assume that the light fields thermalize quickly and give rise to a radiation bath of temperature $T$ and energy density $\rhoR$ whose evolution is described by
\be
\dot \rhoR + 4 H \rhoR = \Ups \dot \phi^2\,. \label{eq:eomRhoR}
\eeq
The dissipation coefficient $\Ups$ parameterizes the energy transfer from the inflaton to the radiation bath and hence appears as an additional friction term in the inflaton's equation of motion,
\be
\ddot \phi + (3H + \Upsilon) \dot\phi + V_\phi = 0\,. \label{eq:eomPhi}
\eeq

In the early days of warm inflation, it was assumed that the production of low-momentum, off-shell $\chi$ particles dominates dissipation since on-shell $\chi$ production is suppressed by the Boltzmann factor $\ee^{-\mchi/T}$. Later on, however, it was realized that, for sufficiently small values of $h$ and $\mchi/T$, on-shell particle production near the pole of the spectral density can be the dominant mode of dissipation after all~\cite{BasteroGil:2012cm}.

In this work, we take into account both the pole and the low-momentum contributions, leading to the general expression
\be
\Upsilon = \Upsilon_{\rm LM} + \Upsilon_{\rm pole}\label{eq:UpsFull}
% \CphiPole \sqrt{\mchi T}\, \ee^{-\mchi/T} + \CphiLM \, \frac{T^3}{\mchi^2} \\
\equiv \Cphi \, \frac{T^3}{\phi^2},
\eeq
with
\begin{align}
\Upsilon_{\rm LM} &= \CphiLM \, \frac{T^3}{\mchi^2}, \\
\Upsilon_{\rm pole} & = \CphiPole \sqrt{\mchi T}\, \ee^{-\mchi/T},
\end{align}
where $\mchi=g \phi/\sqrt{2}$. The various dissipation coefficients
are given by 
\begin{align}
\CphiPole &= \frac{32}{\sqrt{2 \pi}} \frac{g^2 N_\chi}{h^2 N_Y}\,, \\
\CphiLM &= 0.01\, h^2 N_Y\, g^2 N_\chi\,, \\
\Cphi &= \frac2{g^2} \left(\CphiPole \left( \frac{\mchi}{T} \right)^{5/2} \ee^{-\mchi/T} + \CphiLM\right)\,,
\end{align}
where $N_\chi$ and $N_Y$ are the multiplicities of the $X$ and $Y$
fields, respectively. 

\begin{figure}[ht]
\centering
\includegraphics[width=0.5\textwidth]{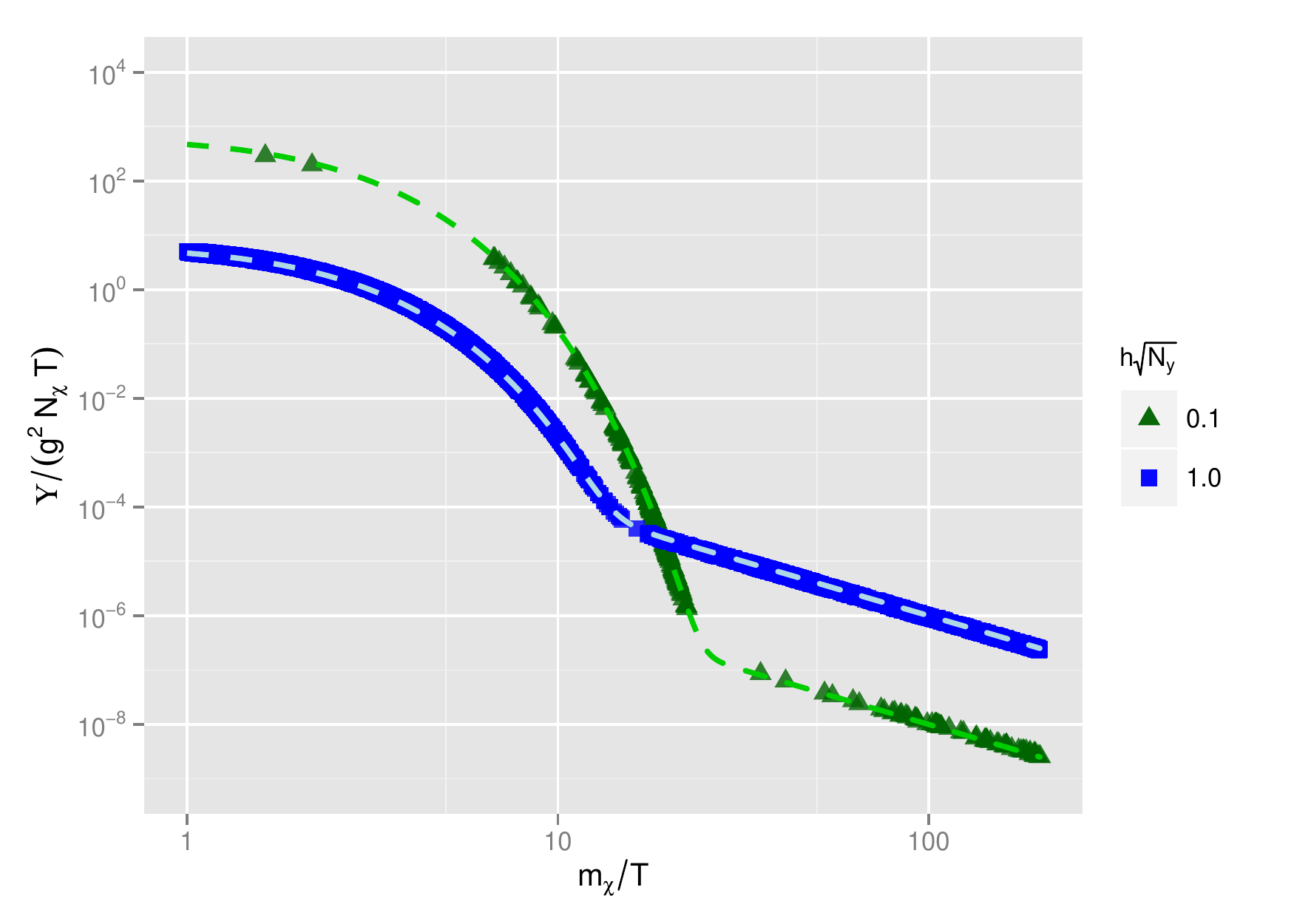}
\caption{Full dissipation coefficient as a function of $\mchi/T$ for
  effective couplings $\hat h = h \sqrt{N_Y} = \{0.1,1.0\}$. Compare
  fig.~(10) of~\cite{BasteroGil:2012cm}. The dashed lines represent the numerical
  prediction made there (cf. their eq.~(4.17)). The data points have been obtained from the simulations presented in this work: green triangles stand for points with $h=0.1$, blue squares $h=1$.}
\label{fig:UpsilonT_mchiT}
\end{figure}

In the low-temperature regime $(T<\mchi)$ we are considering here, the
pole term in eq.~\eqref{eq:UpsFull} dominates for $\mchi/T \sim
\curlO(1)$. In this regime, the sharp peak in on-shell
$\chi$~production more than compensates for the Boltzmann suppression,
resulting in the enhanced dissipation seen in
fig.~\eqref{fig:UpsilonT_mchiT}. For $\mchi/T \gtrsim 15$, the
exponential suppression of the pole term allows the low-momentum term
to dominate once again. The main purpose of this work is to show that
the enhancement of dissipation in the pole regime allows for a significant era
of warm inflation with smaller $N_\chi$ and $g$ than the low-momentum
regime. 

\section{Observables}\label{sec:Observables}

%%\subsection{Spectral index}

For $Q  \lesssim 0.1$, quantum and thermal perturbations lead to a perturbation amplitude given by \cite{Bartrum13,BGBMR14} % assuming negligible inflaton occupation number n_\star at horizon crossing
\be
\PR \simeq \left( \frac{H_\star}{2\pi} \right)^{\!2} \left( \frac{H_\star}{\dot\phi_\star} \right)^{\!2} \left( 1 + \frac{T_\star}{H_\star} \frac{2\pi Q_\star}{\sqrt{1 + \frac{4\pi}{3} Q_\star}} \right) \label{eq:PR}
\eeq
The spectral index is given by
\be
\ns-1 = \frac{\rd \ln P_{\mathcal R}}{\rd\ln k} \simeq  \frac{\rd \ln P_{\mathcal R}}{\rd \Ne}\,,
\eeq
where $\Ne$ is the number of e-folds, and which leads to
\begin{equation}
\begin{aligned}
\ns-1 &= \frac{\epsilon_\phi}{1+Q_\star}  \left(-6 + \frac3{2} \frac{\Delta_\star}{1 + \Delta_\star} + 
\left( 2Q_\star + A  \frac{\Delta_\star}{1 + \Delta_\star} \right)  \frac{2
  + \ceff}{4-\ceff + Q_\star(4 + \ceff)} \right)\\
       &\hphantom{=}+ \frac{\eta_\phi}{1+Q_\star} \left(2 - \frac12
\frac{\Delta_\star}{1 + \Delta_\star}  - 
\left( 2 Q_\star  + A  \frac{\Delta_\star}{1 + \Delta_\star} \right) \frac{2 \ceff}{4-\ceff + Q_\star(4 + \ceff)} \right) \\
             &\hphantom{=} - \frac{\sigma_\phi}{1+Q_\star}  \left( 2 Q_\star  + A  \frac{\Delta_\star}{1 + \Delta_\star} \right) \frac{4(1 - \ceff)}{4-\ceff + Q_\star(4 + \ceff)}\,,
\end{aligned}
\label{eq:ns1}
\end{equation}
where all quantities are evaluated at horizon crossing, denoted by a ``$\star$''. We have used
the slow-roll parameters 
\be
\epsilon_\phi = \frac{\mP^2}{2} \left( \frac{V_\phi}{V} \right)^{\!2}\,, \qquad
\eta_\phi = \mP^2 \frac{V_{\phi\phi}}{V}\,, \qquad
\sigma_\phi = \mP^2 \frac{V_{\phi}}{\phi V}\,,
\eeq
and defined
\begin{align}
\ceff &= \frac{3\Upsilon_{\rm LM}}{\Upsilon} + \frac{\Upsilon_{\rm pole}}{\Upsilon} \left(\frac12
  + \frac{\mchi}{T}\right)\,, \\
\Delta_\star&=  \frac{T_\star}{H_\star} \frac{2\pi Q_\star}{\sqrt{1 +
    \frac{4\pi}{3} Q_\star}} \,, \\
A &= \frac{15 + Q_\star(9 + 12\pi + 4\pi Q_\star)}{4(3+4\pi Q_\star)}\,.  
\end{align}
The above analytical expressions for the amplitude of the spectrum and
the spectral index hold only in the weak dissipative regime, $Q_\star
\lesssim 0.1$. For
larger values of the dissipative coefficient, the radiation
fluctuations backreact onto the field fluctuations, inducing an enhancement of
the spectrum \cite{Graham:2009bf,BasteroGil:2011xd,BGBMR14} that has
to be computed numerically; we are not going to explore that regime in this work.
In the limit that dissipation at horizon crossing is dominated by the
low-momentum modes, with $\ceff \simeq 3$, we recover previous
expressions for the spectral index given in the literature
\cite{Bartrum13,BGBMR14}. In the limit of very weak
dissipation $Q_\star \ll 1$ and 
$\Delta_\star \ll 1$, we just recover the standard cold-inflation
expression for the spectral index:
\beq
\ns-1 \simeq - 6 \epsilon_\phi + 2 \eta_\phi \,.
\eeq
However, this does not mean that predictions are the same as in
cold inflation. Even if starting inflation with a small amount of
dissipation, the dynamics can increase the value of $Q$, which in
turns affects inflaton evolution and therefore
the values of the slow-roll parameters. Typically, when $Q$ increases,
due to the extra friction added by the dissipation, smaller values of
the field are required in order to get $\Ne \sim \curlO(60-50)$. 

\begin{figure}[ht]
\centering
\begin{tabular}{cc}
\!\!\!\!\!\!\includegraphics[height=0.4\textwidth]{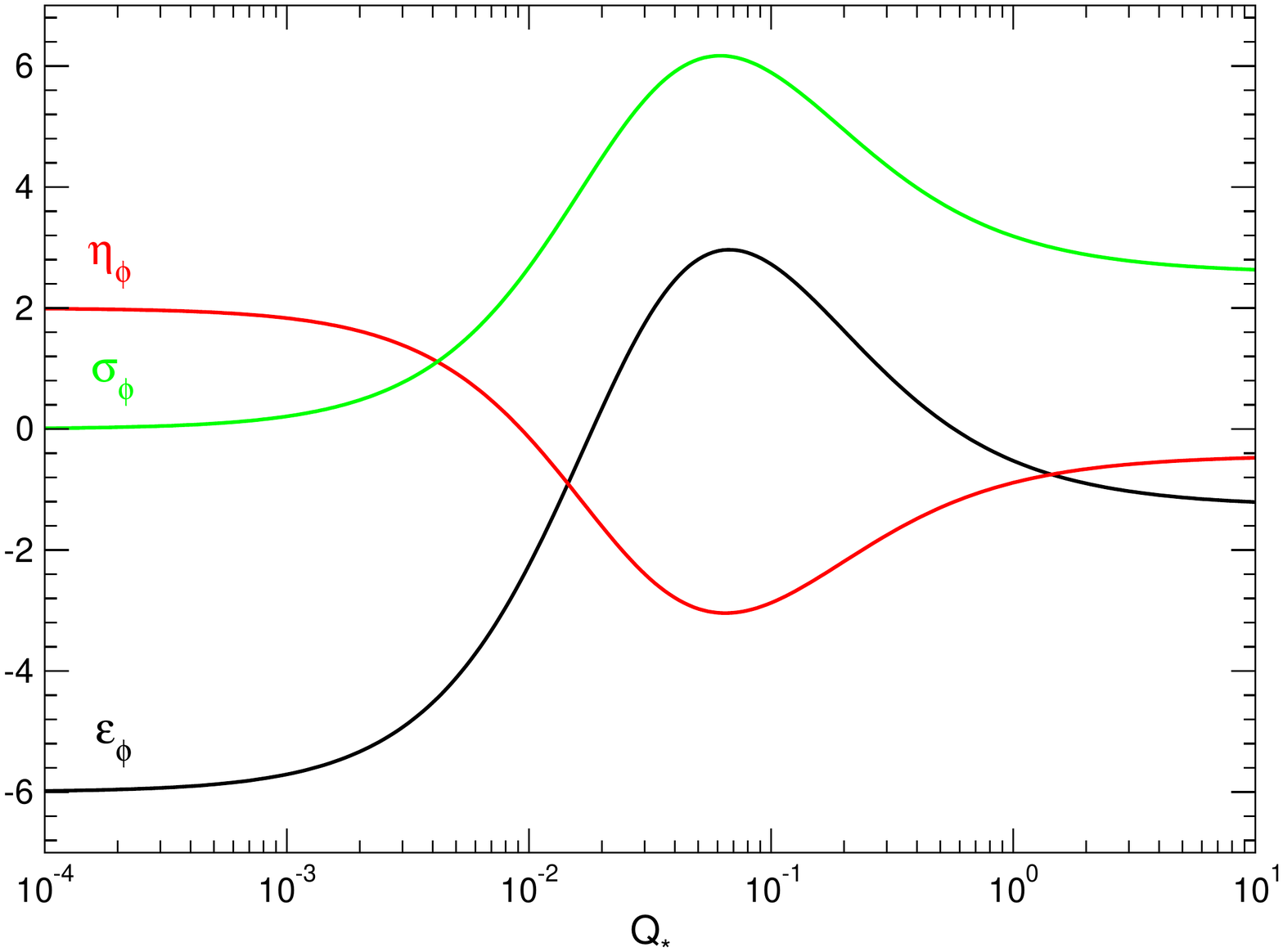}
& 
\!\!\!\!\!\!\includegraphics[height=0.4\textwidth]{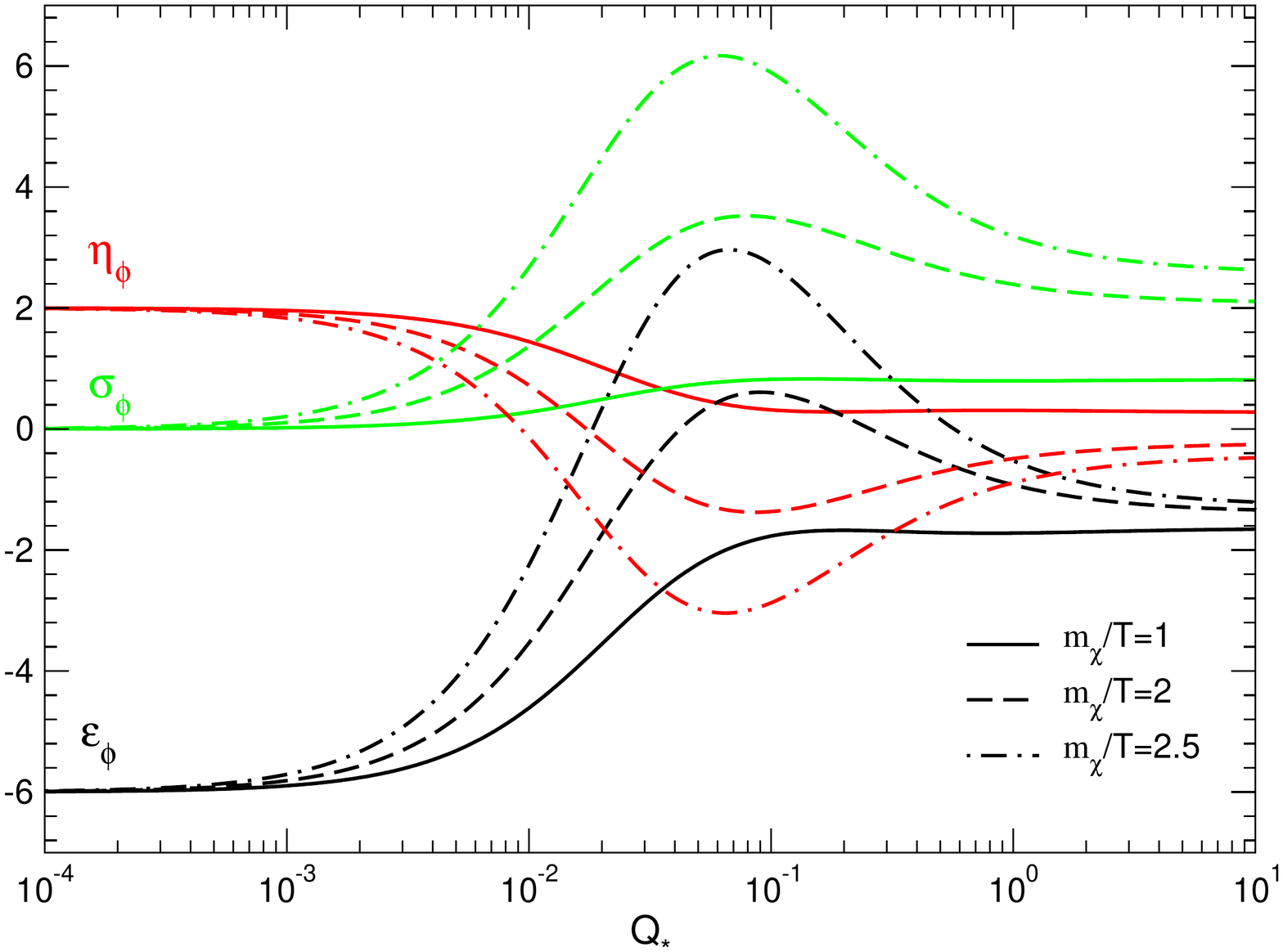}
\end{tabular}
\caption{Left: Coefficients of the slow-roll parameters $\epsilon$,
  $\eta$, and $\sigma$ in the low-momentum limit of
  equation~\eqref{eq:ns1} for the spectral index, i.e., $\ceff =
  3$. Right: Same in the pole-dominated regime,
  $\Upsilon \simeq \Ups_{\rm pole}$, for different values of
  $\mchi/T$. For small $Q$, the coefficients take their cold-inflation
  values $\{-6,2,0\}$. } 
\label{fig:ns1Coeffs_Q}
\end{figure}

For $Q\gtrsim10^{-3},$ the form of the spectral index~\eqref{eq:ns1}
changes, with the coefficients now functions of $Q$, as illustrated in
fig.~\eqref{fig:ns1Coeffs_Q}. In the left panel we have plotted the
coefficients for low-momentum-dominated dissipation, while in the
right panel we have the pole-dominated case for different
values of $m_\chi/T=1,\,2,\,2.5$, i.e., different values of
$\ceff=1.5,\, 2.5,\, 3$. At large $Q$, the coefficients decrease as
  $1/Q$.  

The coefficients depend also on the combination
$Q_*(T_*/H_*)$. During slow roll, this quantity can be derived from the
radiation equation~\eqref{eq:eomRhoR} and the perturbation spectrum~\eqref{eq:PR}
\be 
\frac{T_\star}{H_\star} = \left[
  \frac{45}{8\pi^4}\, \frac{Q_\star}{g_\star \PR}\! \left(\! 1 +
  \frac{T_\star}{H_\star} \frac{2\pi Q_\star}{\sqrt{1 + \frac{4\pi}{3}
      Q_\star}}\! \right) \right]^{1/4}\!, \label{eq:THfromPR} 
\eeq
Indeed, we can use the Planck observation $\PR= 2.2 \times 10^{-9}$ to put a first constraint on the amount of dissipation required for warm inflation from eq.~\eqref{eq:THfromPR}. We conclude that for $g_\star=2$ the warm-inflation condition $T>H$ can be satisfied as long as $Q_\star > 8 \times 10^{-8}$, showing that even a very small amount of dissipation can be enough to produce an era of warm inflation.

The primordial tensor perturbation in warm inflation is given by its
standard vacuum form:
\be
P_T = 8\left(\frac{H_\star}{2 \pi \mP} \right)^2 \,,
\eeq
but the tensor-to-scalar ratio gets modified due to the thermal
contribution to the scalar spectrum, 
\be
r = \frac{16 \epsilon_\star}{(1 + Q_\star) (1 + \Delta_\star)}\,,
\label{tensortoscalar}
\eeq
where $\epsilon_\star= \epsilon_\phi/(1+Q)$.

\section{Potentials with $\mathbf{m_{\boldsymbol\chi}/T=}\text{const.}$}\label{sec:mchiTconst}

We are mainly interested in exploring the possibility of warm
inflation in the pole-dominated regime. Although there is clearly an
enhancement of the dissipative coefficient compared to the low-momentum for 
$m_\chi/T \simeq \curlO(1-10)$, as seen in
fig.~\eqref{fig:UpsilonT_mchiT}, the dissipative coefficient is
suppressed by the Boltzmann factor $\ee^{-m_\chi/T}$. Whenever the
ratio $m_\chi/T$ increases during inflation, the pole contribution may
quickly vanish, so we first explore which kind of potentials may
render this ratio approximately constant during slow-roll inflation. 

We  derive an equation of motion for $x:=\phi/T$ in warm
inflation, starting from
% \begin{widetext}
\be \frac{x^\prime}{x} = \frac{\phi^\prime}{\phi} - \frac{T^\prime}{T}
= \frac{\phi^\prime}{\phi} - \frac14\frac{\rhoR^\prime}{\rhoR} \,,  
\eeq
% \end{widetext}
where a ``prime'' denotes derivative with respect to the number of
e-folds. During slow roll, the energy density in radiation is given by 
\be
\rhoR = \frac34 \frac{Q}{(1+Q)^2} \frac{V_\phi^2}{3H^2}\,, 
\eeq 
From this, we obtain
\be 
\frac{\rhoR^\prime}{\rhoR} = \frac{1-Q}{1+Q} \frac{Q^\prime}{Q} +
2\frac{V_\phi^\prime}{V_\phi} - 2\frac{H^\prime}{H}\,.  \eeq
From the definition of the dissipative ratio, $Q = \Upsilon/(3H)$, 
with $\Upsilon$ given in eq.~\eqref{eq:UpsFull}, 
%\be \Upsilon = \Upsilon^{\rm LM} + \Upsilon^{\rm pole} = \CphiLM
%\frac1{a^2 g^2} \frac{\phi}{x^3} + \CphiPole \frac{\phi}{\sqrt{x}}
%\ee^{-ag\phi/T}\,, 
%\eeq 
we find 
\be \frac{Q^\prime}{Q} = -
\frac{H^\prime}{H} + \frac{\phi^\prime}{\phi} - \ceff
\frac{x^\prime}{x}\,.  
\eeq
This yields the equation of motion for $x$, 
\be 
\frac{x^\prime}{x} = \frac1{4 -\ceff+
  Q(4 + \ceff)}
\left( -\frac{3+Q}{1+Q}\, \epsilon_\phi + 2\eta_\phi -
\frac{3+5Q}{1+Q}\, \sigma_\phi \right)\,.\label{eq:dx} 
\eeq 
Hence, we determine potentials that
exhibit constant $\phi/T$ by setting $x^\prime=0$ and integrating
twice the resulting relation between the potential and its
derivatives, 
\be \frac{V_{\phi\phi}}{V_\phi} - 
\frac{3+Q}{4(1+Q)}\, \frac{V_\phi}{V} = \frac{3+5Q}{1+Q}\,
\frac1{2\phi}\,.  
\eeq 
For $Q\gg1$, this yields a potential 
\be
V^{Q\gg1} = \ee^{C_1} \left( C_2 + \phi^{7/2} \right)^{\!4/3}\, 
\eeq
where $C_1$ and $C_2$ are integration constants. For $Q\ll1$, we get
\be 
V^{Q\ll1} = \ee^{C_1} \left( C_2 + \phi^{5/2} \right)^{\!4}\,.
\eeq 
Depending on the whether $\phi$ is super- or sub-Planckian, we
can write these as either chaotic or hybrid potentials: for
$\phi>\mP$, 
\be 
V^{Q\gg1} \approx V_0
\left(\frac{\phi}{\mP}\right)^{14/3}\,,\qquad V^{Q\ll1} \approx V_0
\left(\frac{\phi}{\mP}\right)^{10}\,, 
\eeq 
and for $\phi<\mP$, 
\be
V^{Q\gg1} \approx V_0 \left( 1 + \tilde \gamma
\left(\frac{\phi}{\mP}\right)^{\!14/3} \right)\,,\qquad V^{Q\ll1}
\approx V_0 \left( 1 + \tilde \gamma \left(\frac{\phi}{\mP}\right)^{\!10}
\right)\,,
\eeq 
where we have defined $V_0=\lambda \mP^4$ for monomial
and $V_0=\lambda \phic^4$ for hybrid potentials, $\phic$ being the
critical field value at which we expect inflation to end via the
waterfall transition.

On top of chaotic and hybrid potentials, we will also study hilltop
potentials, for different powers of the field. The potentials are
then: 
\be
\begin{aligned}
\text{Chaotic:}\; &V = V_0 \left(\frac{\phi}{\mP}\right)^p \,,\\
\text{Hybrid:}\;  &V = V_0 \left( 1 + \frac{\gamma}{p} \left(\frac{\phi}{\mP}\right)^p \right)\,,\\
\text{Hilltop:}\; &V = V_0 \left( 1 - \frac{\gamma}{p} \left(\frac{\phi}{\mP}\right)^p \right)\,.\\
\end{aligned}
\eeq

\section{Numerical Algorithm}\label{sec:Numerics}

To scan the parameter space of our models for points that allow for a
significant amount of warm inflation, we first find initial conditions
near a slow-roll trajectory. Once we have identified suitable initial
conditions, we check whether they satisfy the necessary constraints
for warm inflation. If they do, we let the system evolve until either
the slow-roll or the warm-inflation conditions break down or until
radiation dominates over the inflaton's potential energy. The
conditions we must verify at each stage for the analytical
calculation of the dissipative coefficient eq.~\eqref{eq:UpsFull} to
hold are: $T \geq H$, $m_\chi \geq T$, and the adiabaticity
condition on the decay rate of the $\chi$ fields $\Gamma_\chi \geq
H$. 

If $T>H$, we can translate the requirement that the system be in the
low-temperature regime, $\frac{\mchi}{T} = \frac{\mchi}{H} \frac{H}{T}
> 1$, into the necessary (but not sufficient) condition $\mchi/H >
1$ or 
\be 
g > \frac{\sqrt{V}}{\phi\, \mP}\,.  
\eeq 
For the potential
$V=\lambda \phi^4$ with $\phi\sim \curlO(1)$ and
$\lambda\sim\curlO(10^{-14}),$ for instance, this translates to
$g>10^{-7}$. In fact, we can tighten this constraint by requiring
adiabaticity, $\Gamchi > H$, which yields 
\be 
g > \frac{64\pi}{h^2  N_Y} \frac{\sqrt{V}}{\phi\,\mP}\,.  
\eeq
While small values of $g$ and $h$ favor the pole-dominated regime we are interested in, the above consistency constraints show that the coupling constants cannot be lowered arbitrarily while keeping particle production strong enough for warm inflation and ensuring that the particles produced thermalize quickly.

In order to scan the parameter space, we begin by generating random values for the coupling constants $g$
and $h$, the initial values of $\phi$ and $\rhoR/V_0$, the number $\Nchi$ of
mediator fields, and in the case of hybrid and hilltop
potentials, the coupling constant $\gamma$. We then obtain slow-roll initial conditions by
simultaneously solving the equation for the Hubble parameter and the
slow-roll versions of the equations \eqref{eq:eomRhoR} and
\eqref{eq:eomPhi} for $V_0$, $\dot\phi$, and $H$. Hence, the
relevant equations are,\footnote{In order to ensure non-negativity of
  $V_0$ and $H$ and to avoid some of the possible numerical problems,
  we work with the logarithms of these equation.}
\begin{align}
V_0 &= (3H + \Upsilon)\,\frac{-\dot\phi}{v_\phi}\,, \\
-\dot\phi &= \sqrt{\frac{4H\rhoR}{\Upsilon}}\,, \\
H &= \sqrt{\frac{V_0 v + \rhoR + \frac12 \dot\phi^2}{3\mP^2}}\,,
\end{align}
where we have defined $v = V/V_0$. We then use $V_0$ to fix one final parameter for each model in order to ensure slow-roll conditions: for monomial and hilltop potentials, we obtain the coupling constant $\lambda$ from $V_0 = \lambda \mP^4$; for hybrid potentials, we set $\lambda = g^2$ and use $V_0 = \lambda \phic^4$ to fix the critical field value for the waterfall transition in these models.
%  and finally arrive at the set of equations
% \begin{align}
% \ln V_0 &= \ln(3H + \Upsilon) + \ln\left(\frac{-\dot\phi}{v_\phi} \right) \\
% \ln\left|\dot\phi\right| &= \frac12 \left(\ln(4H) + \ln\rhoR - \ln\Upsilon\right) \\
% \ln H &= \frac12 \ln\left(V_0 v + \rhoR + \frac12 \dot\phi^2\right) - \ln\left(3\mP^2\right)\,,
% \end{align}

Given these initial conditions, the system should find itself close to
a slow-roll trajectory. We proceed by integrating numerically the full
equations of motion for the inflaton, the radiation density, and the
scale factor,
\begin{align}
\frac{\rd^2\phi}{\rd t^2} &= -\frac{\rd V}{\rd\phi} - (\Upsilon+3H)
\frac{\rd \phi}{\rd t}\,,\\ \frac{\rd \ln \rhoR}{\rd t} &=
\frac{\Upsilon}{\rhoR} \left(\frac{\rd \phi}{\rd t}\right)^{\!2} -
4H\,,\\ \frac{\rd \ln a}{\rd t} &= H\,.
\end{align}
Evolution ends when the slow-roll conditions or the conditions for
warm inflation break down. We keep any parameter points that produce at least 1~$\ee$-fold of inflation and compare to observational data those that lie between 45 and 55 $\ee$-folds.

\section{Results}\label{sec:Results}

\begin{figure}[ht]
\centering
\includegraphics[width=\textwidth]{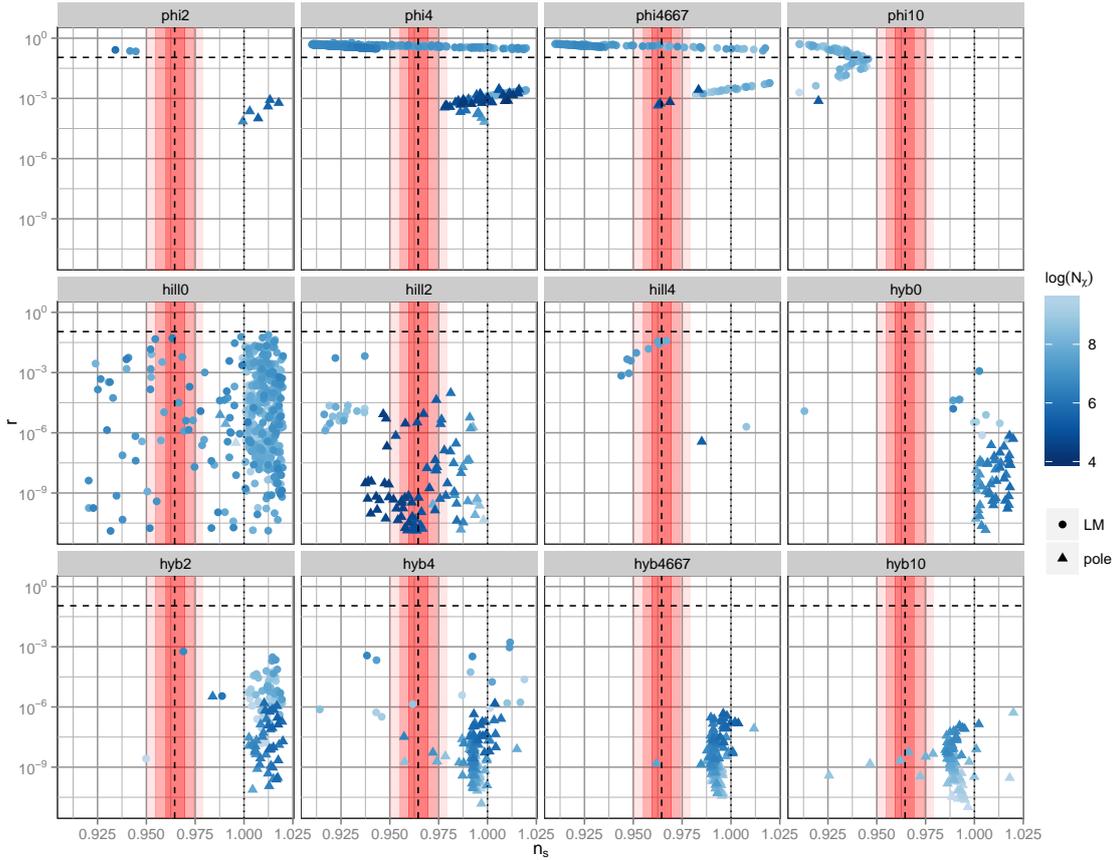}
\caption{Tensor-to-scalar ratio $r$ vs spectral index $\ns$ for
  monomial, hilltop, and hybrid potentials. Triangles represent
  pole-dominated, disks low-momentum-dominated points; color
  represents the number of mediator fields, $\Nchi$. All points lie
  between 45 and 55 $\ee$-folds. The dashed black line and shaded
  intervals indicate, respectively, the central value and $1\sigma$,
  $2\sigma$, and $3\sigma$ confidence intervals of $\ns$ based on the
  Planck+lensing data \cite{Planck13Inflation}.}
\label{fig:ns_r}
\end{figure}

In order to assess the viability of our models, we compare our
predictions for the spectral index and the tensor-to-scalar ratio to
observations by the Planck satellite~\cite{Planck15Params} in fig.~\eqref{fig:ns_r}. We show the
results for chaotic (phi$p$), hilltop (hill$p$), and hybrid (hyb$p$)
potentials, for different powers of the field $p$ as indicated in the
figure (the label ``$4667$'' refers to $p=14/3$). For hilltop and hybrid, $p=0$ refers to a logarithmic potential:
\be
V = V_0 \left( 1 \pm \gamma \ln \left(\frac{\phi}{\mP}\right) \right)\,. 
\eeq

\begin{figure}[ht]
\centering
\includegraphics[width=\textwidth]{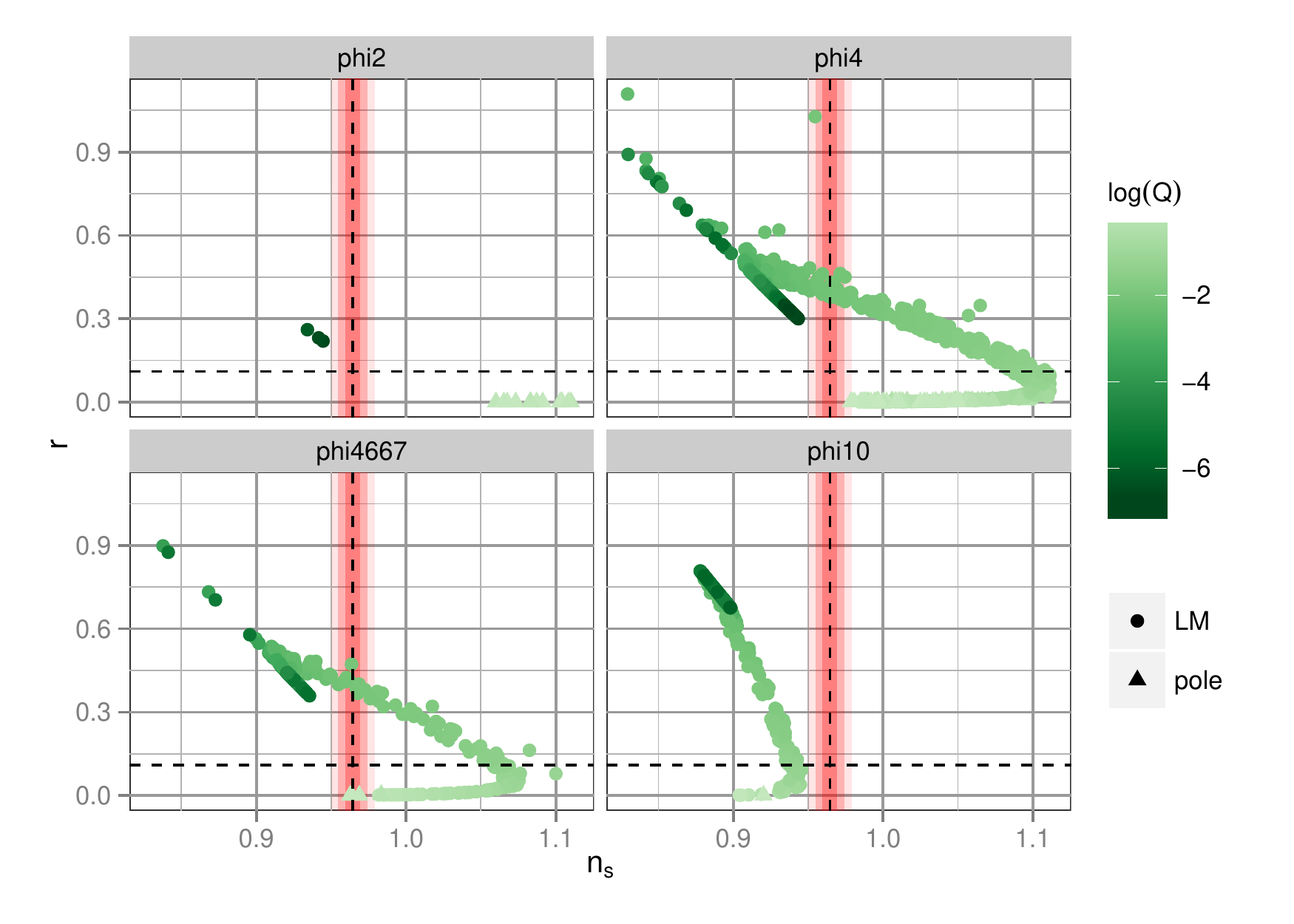}
\caption{Tensor-to-scalar ratio $r$ vs spectral index $\ns$ for
  monomial potentials with exponents
  $p={2,4,\frac{14}{3},10}$. Triangles represent pole-dominated, disks
  low-momentum-dominated points; color represents the dissipative
  ratio, $Q$. All points lie between 45 and 55 $\ee$-folds. The dashed
  black line and shaded intervals indicate, respectively, the central
  value and $1\sigma$, $2\sigma$, and $3\sigma$ confidence intervals
  of $\ns$ based on the Planck+lensing data
  \cite{Planck13Inflation}.}
\label{fig:ns_r_linear}
\end{figure}

For monomial potentials, fig.~\eqref{fig:ns_r_linear} shows the $\ns$--$r$ plane separately with a linear $r$ axis to emphasize the large-$r$ region. We can see that, for increasing $Q$, the trajectory in that plane follows an arc similar to the one seen in refs.~\cite{Bartrum13,BasteroGil:2014oga}. We find low-momentum-dominated points at low $Q$ that
allow for a spectral index compatible with Planck results for the
$\phi^4$ and $\phi^{14/3}$ models; these points do, however, have
tensor-to-scalar ratios much bigger than the Planck constraint
$r<0.11$. At larger $Q$ and smaller $r$, the trajectory for these two
potentials returns to the Planck range for the spectral index; those
points tend to be pole-dominated and have $r<10^{-3}$ (compare
fig.~\eqref{fig:ns_r}). We observe further that the maximum $\ns$
for each of these models is reached around $Q\approx 5\times 10^{-2}$
and $T/H\approx 50$; this large-$\ns$ cusp of the trajectory moves
to smaller and smaller $\ns$ as the exponent of the potential
increases. At its largest values, $\ns$ is dominated by the (positive)
contributions from the $\eta$ and $\sigma$ terms in
expression~\eqref{eq:ns1}; the low-momentum points below $Q\approx
5\times10^{-3}$ and $T/H \approx 20$, where the coefficients of
$\epsilon$ and $\eta$ go through zero, are dominated by the (negative)
contribution of the $\epsilon$ term.

For hybrid potentials, our data in both the LM and pole regimes tend
to cluster around $\ns=1$, with only the quartic hybrid potential
producing points compatible with Planck data.  The contribution of the
$\epsilon$ term to the spectral index tends to be negligible for our
hybrid data; instead, $\ns$ is set by the (negative) $\eta$ and the
(positive) $\sigma$ term.

Quadratic hilltop potentials, however, show points compatible with
Planck data in the pole regime. As for the hybrid model, $\epsilon$ at
horizon crossing is negligible with respect to $\eta$ and $\sigma$,
and therefore the tensor-to-scalar ratio is suppressed and below $r<
10^{-3}$.     

\begin{figure}[ht]
\centering
\includegraphics[width=\textwidth]{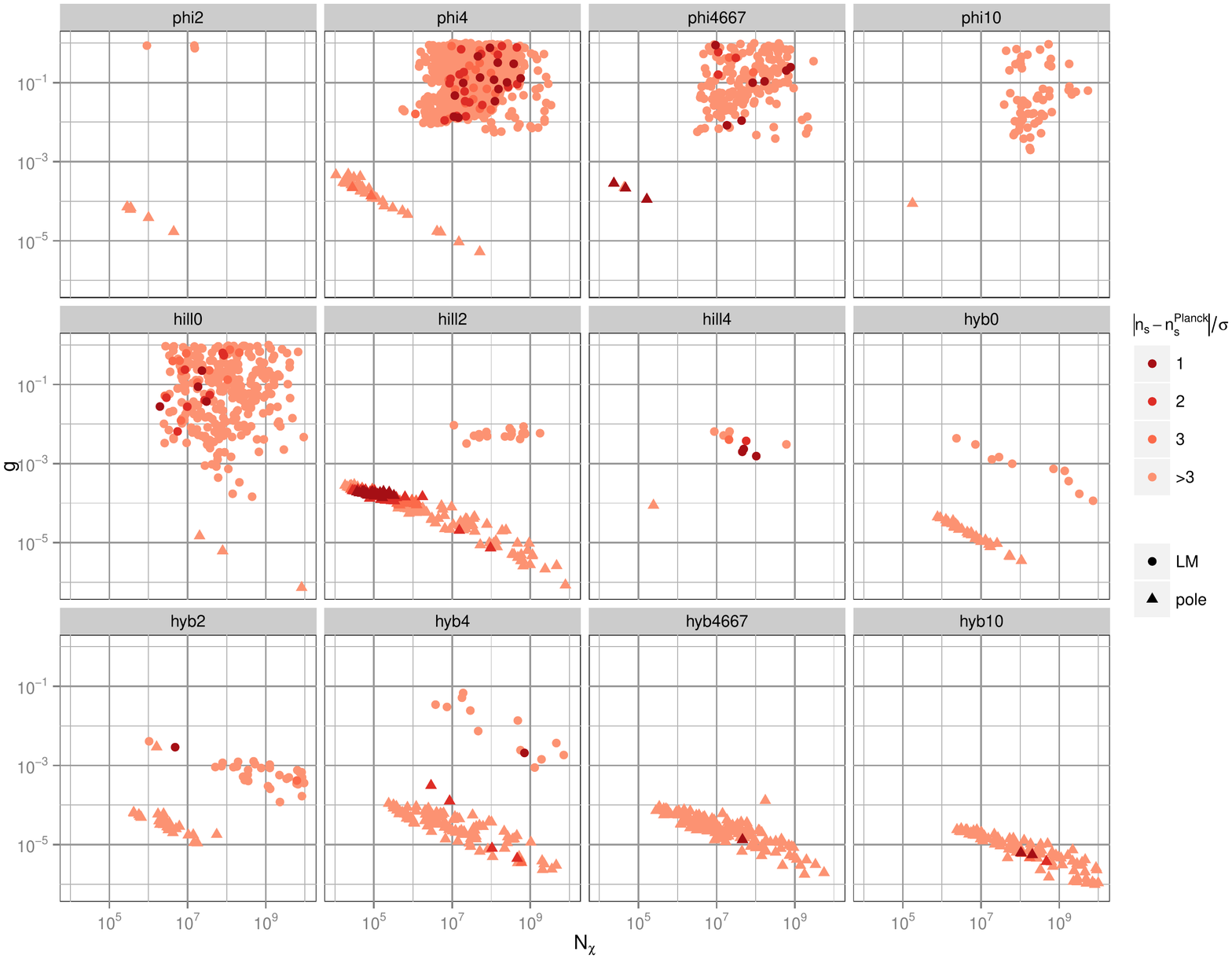}
\caption{Points in the $g$--$\Nchi$ plane that allow for 45--55
  $\ee$-folds of inflation. Color indicates the deviation from the
  central value of $\ns$ as measured by Planck+lensing data
  \cite{Planck13Inflation}. Circles indicate low-momentum-dominated
  dissipation, triangles indicate pole-dominated dissipation.}
\label{fig:Nchi_g}
\end{figure}

\begin{figure}[ht]
\centering
\includegraphics[width=\textwidth]{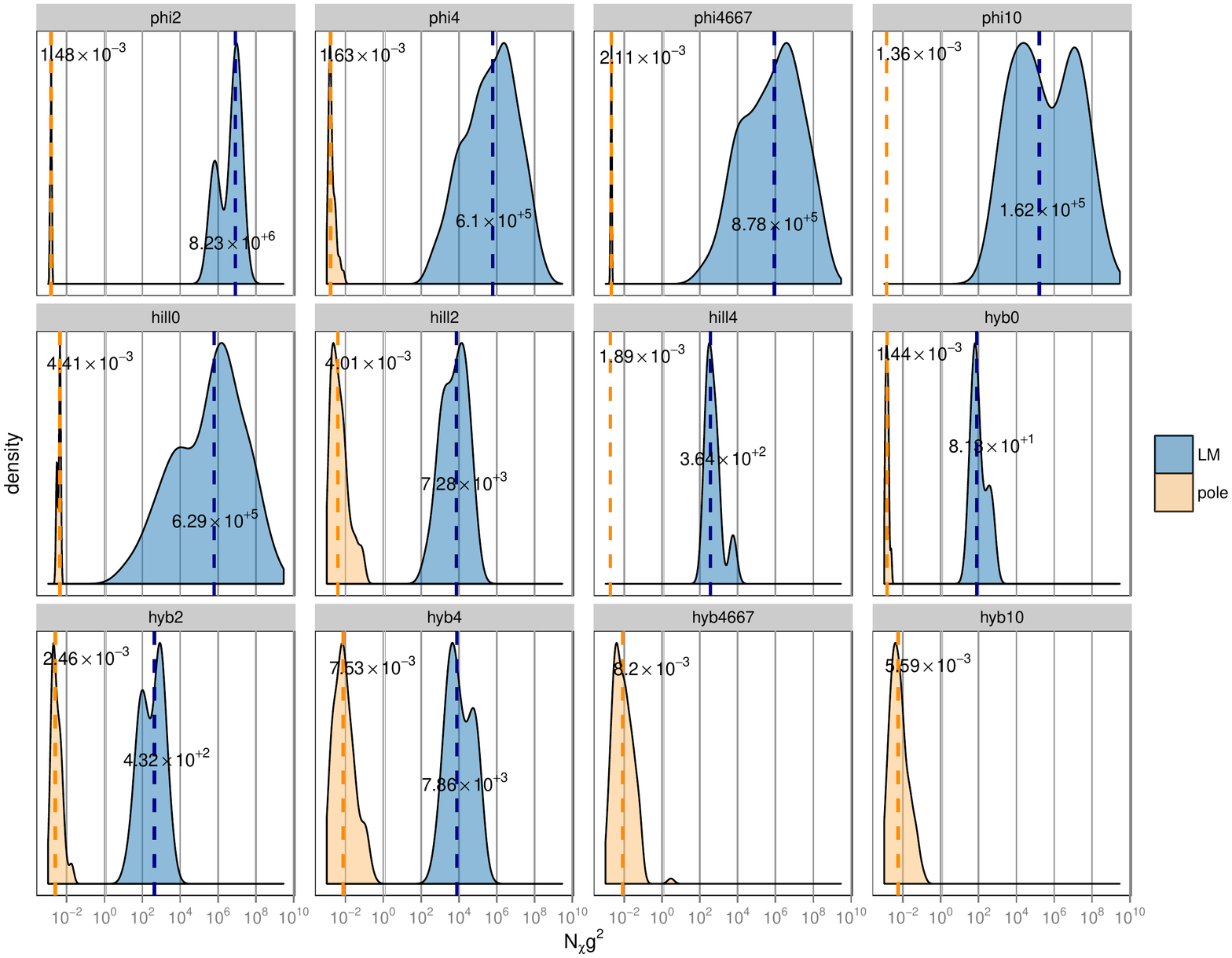}
\caption{Distributions and median values of $\Nchi g^2$ for
  low-momentum- and pole-dominated points between 45 and 55 $\ee$-folds in monomial, hilltop, and hybrid potentials.}
\label{fig:Nchi_g2}
\end{figure}

Figure~\eqref{fig:Nchi_g} illustrates a main advantage of allowing the
pole term to contribute to dissipation:
pole-dominated dissipation allows for a significant amount of warm
inflation with noticeably smaller values of $g$ and $\Nchi$ than the
low-momentum regime does. For all potentials, the pole
and LM regions are cleanly separated, corresponding to the different
ranges of $\mchi/T$ inhabited by the two regimes. The same effect
appears in fig.~\eqref{fig:Nchi_g2}, where we compare $\Nchi g^2$
for low-momentum and pole domination---pole values are consistently
smaller. Once we have picked a value for the coupling $g$ that is
small enough to keep radiative corrections under control,
fig.~\eqref{fig:Nchi_g2} can provide a rough estimate of the number
of mediator fields that need to be introduced to obtain warm
inflation.

\begin{figure}[ht]
\centering
\includegraphics[width=\textwidth]{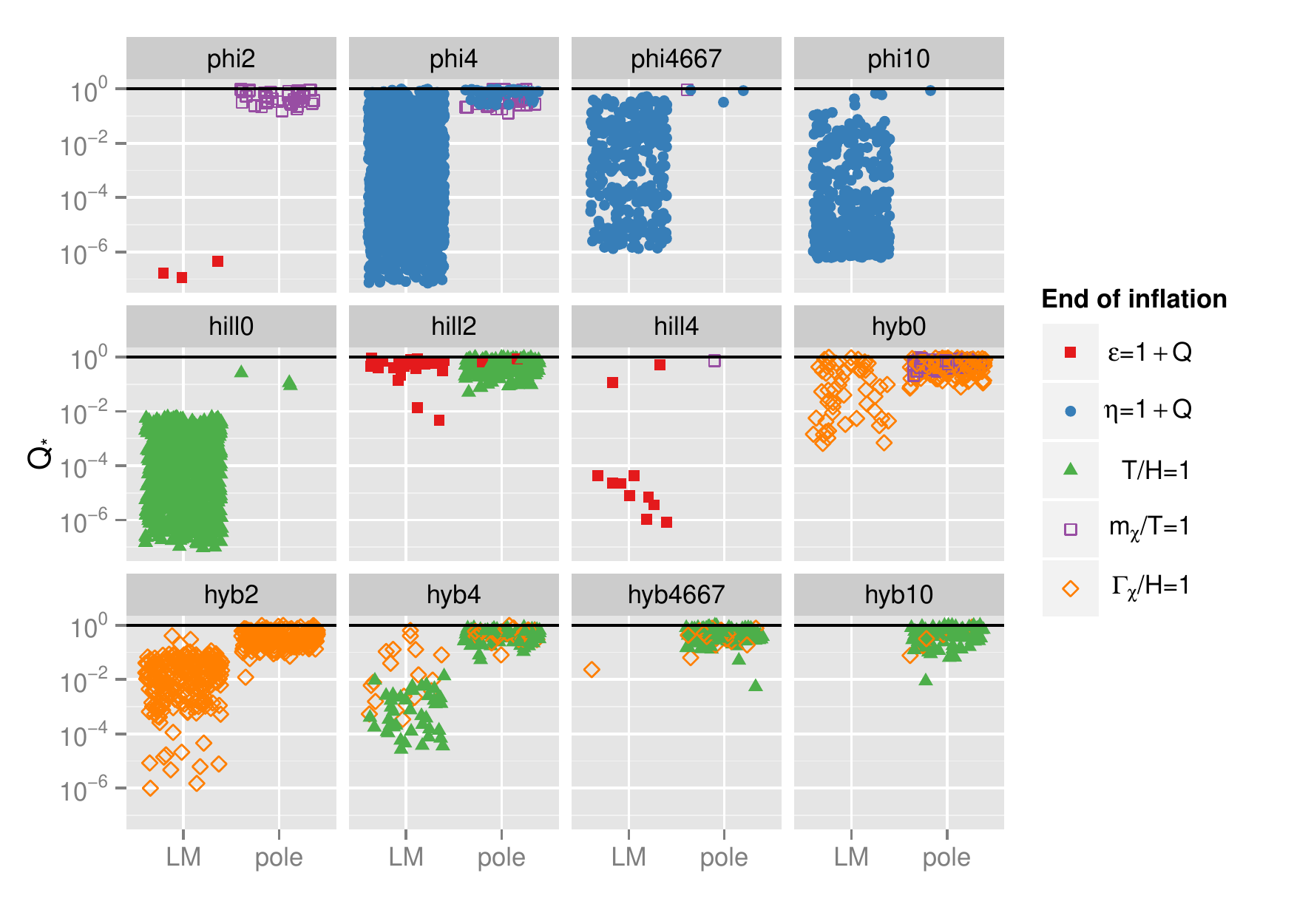}
\caption{Reasons for the end of warm inflation in the pole and LM regimes. For all potentials, the pole regime seems confined to relatively large $Q_\star$. All points lie between 45 and 55 $\ee$-folds and within $10\sigma$ of Planck's spectral index.}
\label{fig:poleLM_Q_rootIndex}
\end{figure}

It is interesting to look at the way warm inflation ends; as shown in
fig.~\eqref{fig:poleLM_Q_rootIndex}, this stopping condition depends
strongly on the potential under consideration. In monomial potentials,
we mostly see a breakdown of slow-roll via $\eta=1+Q$; we will take a
closer look at these points in
section~\eqref{sec:NeBoundForMonomials}. For most of our pole points
in $\phi^2$ and some in $\phi^4$ and $\phi^{14/3}$, warm inflation
ends with $\Gamchi/H=1$.

In the hybrid potentials, $\Gamchi/H=1$ is the dominant mode for
ending warm inflation, but there remain pole points in the logarithmic
and quadratic potentials that end via $\mchi/T=1$; for $p\ge4$, many
points end in $T/H=1$.

Hilltop models, both logarithmic in the LM regime and quadratic in the
pole one, have the parameter $T/H$ decreasing by the end of
inflation and reaching the lower limit $T/H=1$. For the
quadratic model, we have $\eta_\phi=\sigma_\phi <0$, and therefore
from eqs.~\eqref{dTH} and \eqref{dphiT} in
the weak dissipative regime $Q \ll 1$:
\be
\begin{aligned}
\frac{\rd \ln T/H}{\rd \Ne } & \simeq - \frac{3 -\ceff}{4 -\ceff}
\sigma_\phi \,, \\
\frac{\rd \ln \phi/T}{\rd \Ne } & \simeq -\frac{1}{4 -\ceff}
\sigma_\phi \,.
\end{aligned}
\eeq
The ratio $\phi/T$ increases during inflation, and therefore so does
$\ceff$; when $\ceff > 4$, the ratio $T/H$ starts to decrease.

It is worth noting that the end of warm inflation does not imply the end of inflation per se. If the temperature drops below the Hubble rate ($T/H<1$) and dissipation is weak ($Q \ll 1$) but slow-roll persists, warm inflation may be followed by an additional phase of cold inflation. We do not investigate that case any further in this work.

\subsection[{Upper bound on $\Ne$ for monomial potentials}]{Upper bound on $\mathbf{\Ne}$ for monomial potentials}\label{sec:NeBoundForMonomials}

The number of $\ee$-folds of inflation is given by
\be
\Ne = \int_{\phii}^{\phif} \frac{H}{\dot\phi} \rd\phi =
-\int_{\phii}^{\phif} \frac{V}{V_\phi} \frac{1+Q}{\mP^2}
\rd\phi\,. \label{eq:NeFromFieldExcursion}
\eeq
For many of our data points in monomial potentials, inflation ends with
$\eta=1+Q_{\mathrm{f}}$ (cf.~fig.~\eqref{fig:poleLM_Q_rootIndex}),
which fixes the final field value. If we assume constant $Q=Q_{\mathrm
  f}=Q_{\mathrm i},$ we can use the
integral~\eqref{eq:NeFromFieldExcursion} to set an upper limit on the
initial field value for a given number of $\ee$-foldings,
\be
\phii \le \sqrt{\frac{2p \Ne \mP^2}{1+Q_{\mathrm i}} + \phif^2}\,.
\label{eq:phiInitialUpper}
\eeq
Field values below the upper limit are obtained if $Q$ increases over the course of
inflation, which is the case for monomial potentials with exponent
$p<14$ in the low-momentum regime~\cite{BasteroGil09}, and hence for
all monomial potentials considered here. Since the slow-roll
parameters for monomial potentials are functions of $p$ and $1/\phi^2$
only, we can convert this into an upper limit on the low-$Q$ spectral
index $\ns-1 = -6\epsilon + 2\eta$, 
\be 
\ns-1 \le -\frac{2 (p+2)  (1+Q_{\mathrm i})}{p-1 + 4 \Ne}\,. \label{eq:ns1Bound} 
\eeq 
Even for the small values of $Q$ assumed here, dissipation allows $\eta$ to
take greater values without slow roll breaking down, and hence the
final field value is allowed to be smaller than without
dissipation. Additionally, dissipation reduces $\rd\phi/\rd \Ne$, so a
smaller field excursion is necessary to produce a given number of
$\ee$-folds. Dissipation effectively compresses the inflaton field
range and shifts it down to lower field values.

\begin{figure}[ht]
\centering
\includegraphics[width=\textwidth]{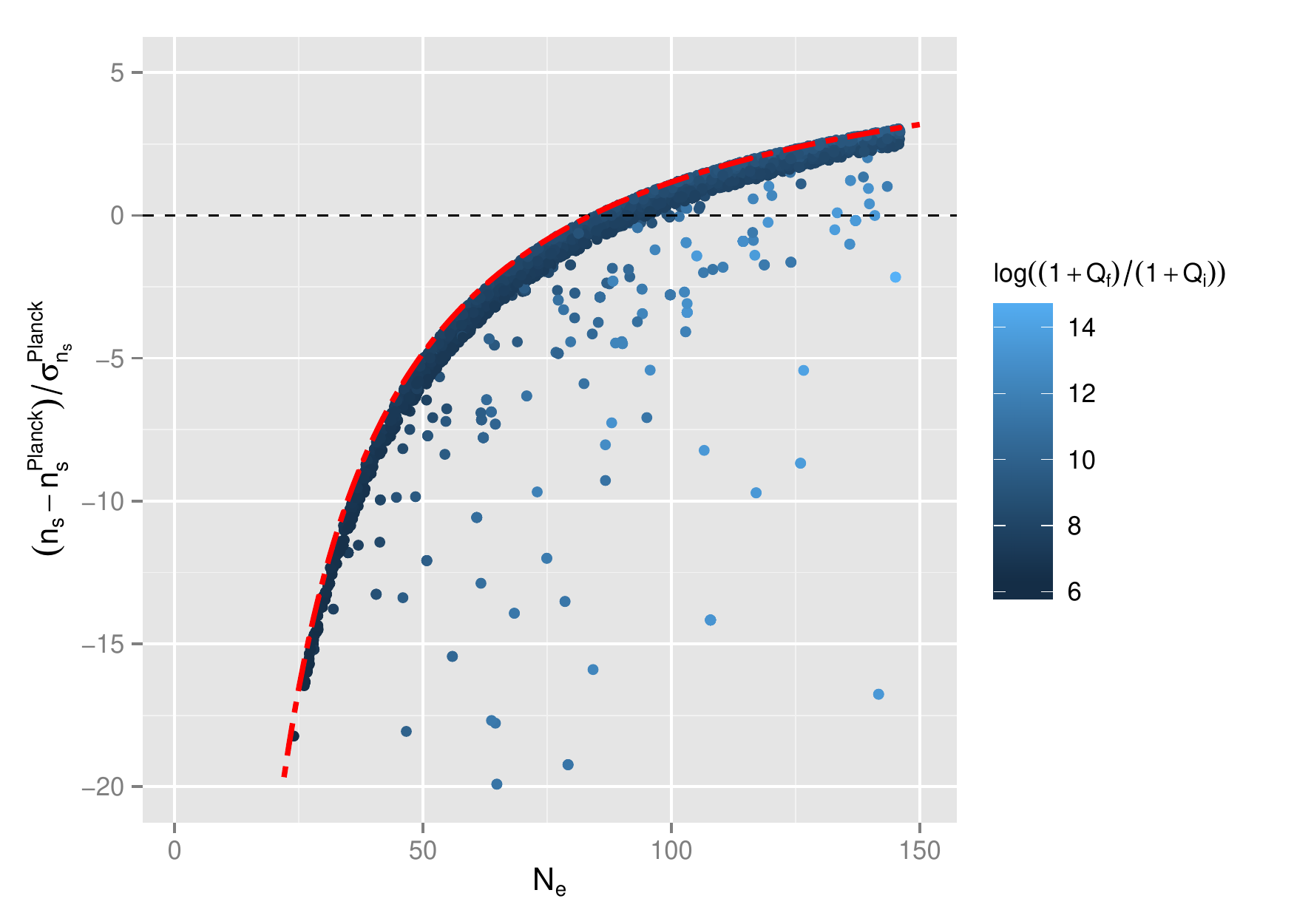}
\caption{Spectral index (in standard deviations from Planck central value) vs $\ee$-folds for low-$Q$, low-momentum data
  in the quartic monomial potential. We have selected points with
  $Q_{\mathrm{i}}<10^{-6}$, where the spectral index has the form
  $\ns-1=-6\epsilon+2\eta.$ The dot-dashed red line indicates the upper
  limit~\eqref{eq:ns1Bound} on $\ns$ if dissipation is negligible and
  inflation ends with $\eta = 1+Q$.}
% ggplot(subset(assignPoleLM(phi4Combined,4,"TH1")[order(1+phi4Combined$QEnd)/(1+phi4Combined$Q4),], poleLM=="LM" & rootIndex==1 & Q4<1e-6)) + geom_point(aes(x=N-4, y=(ns1-ns1_planck)/ns1_planck_sigma, colour=log10((QEnd)/(Q4))
% )) +
%     scale_x_continuous(expression(N[e])) +
%     scale_y_continuous(expression((n[s]-n[s]^{Planck})/sigma[n[s]]^{Planck}),limits=c(-20,5)) +
%     geom_hline(y=0, linetype="dashed") +
%     geom_line(data=data.frame(cbind(Ne=NeSeq,ns1=ns1NeFunction(NeSeq,4,0))),aes(x=Ne, y=(ns1-ns1_planck)/ns1_planck_sigma), colour="red", lwd=1) +
%     scale_colour_continuous(expression(log((1+Q[f])/(1+Q[i]))))
\label{fig:ns_Ne_upperLimit}
\end{figure}

For $Q<10^{-6},$ our data show the expected behavior: the spectral
index at any given $\Ne$ lies below the limit~\eqref{eq:ns1Bound}. For
illustration, we include in fig.~\eqref{fig:ns_Ne_upperLimit} our
low-momentum, low-$Q$ data for the quartic monomial potential
alongside the bound~\eqref{eq:ns1Bound}; the width of the densely-populated band below
the bound is given by the constraint~$Q_{\mathrm{i}}<10^{-6}$ we have
imposed on the data in this plot. Scattered below the band are points
with large $(1+Q_{\mathrm{f}})/(1+Q_{\mathrm{i}})$, for which $Q$
changes dramatically over the course of inflation and has a
significant effect on the integral~\eqref{eq:NeFromFieldExcursion} right from
the start.

It is interesting to note that, for monomial potentials with exponent
$p=\{2,4,\frac{14}{3},10\}$, the low-$Q$ upper limit on $\ns$ enters
Planck's $2\sigma$ range at $\Ne=\{44, 65, 73, 130\},$
respectively---at least this many $\ee$-folds of low-$Q$ warm
inflation in the low-momentum regime are necessary for these
potentials to produce an observationally-viable spectral index.

For $Q\gtrsim10^{-3},$ the form of the spectral index~\eqref{eq:ns1}
changes, with the coefficients now functions of $Q$, as illustrated in
fig.~\eqref{fig:ns1Coeffs_Q}. In this regime, the
bound~\eqref{eq:ns1Bound} no longer applies, and the slow-roll
parameters in the spectral index no longer have their simple
cold-inflation coefficients. For $Q>10$, these coefficients decrease
as $1/Q$ while maintaining fixed ratios ($1:3:-\frac12$) between the
coefficients of $\epsilon$, $\eta$, and $\sigma$.
 % $\eta = \frac13 \epsilon$ and $\sigma = -2\epsilon$.

% \newpage 

% \begin{figure}[ht]
% \centering
% \includegraphics[width=\textwidth]{plots_paper/ns_Ne_Q_3d.pdf}
% \caption{Full shape of the dependence of $\ns$ on $\Ne$ and $Q$ in the low-momentum regime of the quartic monomial potential.}
% \label{fig:ns_Ne_Q_3d}
% \end{figure}

\section{Conclusions}\label{sec:Conclusions}

We have studied warm-inflation dynamics in the
low-$T$ regime with a general dissipation coefficient. Dissipation
originates from the two-stage mechanism \cite{Berera:2001gs,
  Moss:2006gt}, and during inflation 
a small part of the inflaton energy density is transferred, through a heavy mediator with $m_\chi \geq T$, into a thermal bath of light degrees of freedom. In previous analysis, only the low-momentum 
contribution, from off-shell $\chi$  modes,  to the dissipative coefficient was considered when
studying the observational implications of warm inflation
\cite{BasteroGil09, BasteroGil:2011mr, Cerezo:2012ub, Bartrum13}. In
this paper, we have extended the analysis by including on-shell
particle production from $\chi$ modes. Although one expects this
contribution to be Boltzmann-suppressed for a heavy $\chi$ mode, 
 for sufficiently weak couplings the on-shell contribution can dominate
 over the low-momentum one \cite{BasteroGil:2012cm}. 
This is due to the different parametric dependence on the Yukawa
coupling $h$ of the mediator $\chi$ to the light degrees of freedom. 
The key point is that the off-shell
contribution is proportional to $h$, 
whilst the on-shell contribution, which peaks near the
pole of the $\chi$ spectral density, is inversely proportional to
the decay rate of $\chi$ and hence to the coupling $h$. 

Typically, dissipation in the low-momentum regime can sustain a sufficiently long
period of warm inflation consistent with observations, for instance for
the quartic chaotic model. Nevertheless, having enough dissipation
requires a large number of mediator $\chi$ fields in the model. We
wanted to explore the possibility of reducing this large number of
fields by compensating with the enhanced behavior of dissipation in the
pole regime. Thus, we have worked with the most general expression
for the dissipation 
$\Upsilon$ in the low-$T$ regime, eq.~\eqref{eq:UpsFull}, and studied
3 different generic models of single-field inflation: chaotic, hybrid, and
hilltop.
One might assume that, in order to get large $\Upsilon$, it is enough to reduce the value of $h$. Consistency of the approximations when computing the dissipative coefficient, however, requires that $\Gamchi$, the $T=0$ decay rate of $\chi$ into light degrees of freedom, satisfy the adiabaticity condition $\Gamchi>H$; this imposes a lower bound on $h$.
This analysis does, however, suggest
that if $H$ can be lowered while keeping $\Gamchi/h^2$ constant
or decreasing less, the parameter space of warm-inflation solutions
can be extended, a point relevant to consider in future studies.

More generally, this work highlights that, as has been seen in previous warm-inflation model-building work, the need for large field content is
not due to the naive expectation that more fields are needed for more dissipation. Instead, it is the complicated set of constraints warm-inflation models have to satisfy that requires a large number of extra fields. The emerging understanding is that only for an appropriate choice of inflaton potentials and underlying field-theory models will we be able to lower the field content. This paper has demonstrated that point and explicitly shown new ways to reduce the field content by looking at the full parametric dependence of the dissipative coefficient alongside the rest of the model.

The main result on the parameters of the
model is summarized in fig.~\eqref{fig:Nchi_g}. The ``Low Momentum''
and ``pole'' labels refer to which contribution to $\Upsilon$
dominates at the time of horizon crossing, when the CMB observables
are evaluated. There is a clear separation between both regimes
depending on the value of the coupling $g$ between the inflaton and
the mediator: horizon-crossing pole domination requires small values,
$g < 10^{-3}$, in order to keep the ratio~$\mchi/T \lesssim \curlO(10)$ and
avoid large Boltzmann suppression.

It is also clear from fig.~\eqref{fig:Nchi_g} that pole domination allows for smaller numbers of mediator fields than the LM regime. While low-momentum dissipation typically needs a minimum of~$\curlO(10^6)$ fields in any given model, pole-dominated dissipation in quadratic hilltop models, for example, only requires~$\curlO(10^4)$ mediators. Such numbers could be achieved in brane-antibrane models of inflation \cite{Burgess:2001fx}. In fact, it is known that the number of mediators for LM dissipation in brane-antibrane models lies in the range~$\curlO(10^4-10^6)$~\cite{BasteroGil:2011mr}; the present analysis shows that that number might drop when we include the pole regime.
% While typically LM dissipation, for any model, needs minimum $O(10^6)$ fields,  for example  quadratic hilltop models only require $O(10^4)$ mediators in the pole regime. This number of fields could be achieved in brane-antibrane models of inflation \cite{Burgess:2001fx}. Already in the LM dissipative regime the no. of mediators was in the range $O(10^4-10^6)$ \cite{BasteroGil:2011mr}, and the present analyses shows that that number might be reduced when including the Pole regime.

The dynamics can make the ratio $\mchi/T$ increase during inflation, such that after pole domination ends, we can continue in the low-momentum regime. Dissipation may even become negligible, allowing inflation to continue some further
e-folds in the cold regime; this is an open possibility for the
quadratic hilltop model, for example. However, we have not explored those scenarios 
here, and consider only the regime of warm inflation when deriving the
observational constraints. The analysis of the spectrum relies on
analytical expressions that hold only in the weak dissipative regime
$Q_\star < 1$, and that is another restriction imposed on the study. While
we know from previous studies that there is a growing
mode in the LM regime that may lead to too large a tilt in the spectrum \cite{BGBMR14}, such numerical studies have not yet been done for the pole regime. However,
for a $T$-dependent dissipative coefficient $\Upsilon \propto T^c$, we
know that the effect is larger the larger the power $c$
\cite{Graham:2009bf,BasteroGil:2011xd}. With the general dissipative
coefficient, during slow roll it seems that the system behaves
like having a dissipative coefficient with $\Upsilon \propto T^{\ceff}$,
where now $\ceff$ will change during the evolution. In the pole regime,
we have the lower bound $\ceff \geq 3/2$, and therefore radiation fluctuations
may still influence the spectrum. Nevertheless, although it seems
unlikely that the growing mode will disappear completely in the pole
regime, its effect can be diminished.  
   
Our aim was to check the possibility of having warm inflation in the
pole regime, identify the parameter regions, and study the trend of
the observables. It is clear, for example, that the pole regime can give a negligible tensor-to-scalar ratio
even for chaotic models. In general, this new regime opens up new
possibilities for detailed model building with fewer fields.

\appendix 

\section{Slow-roll Equations}

In this appendix we give the slow-roll evolution equations for the parameter $Q$ and the ratios $T/H$ and $\phi/T$, in terms of the slow-roll parameters:
\be
\epsilon_\phi = \frac{\mP^2}{2} \left( \frac{V_\phi}{V} \right)^{\!2}\,, \qquad
\eta_\phi = \mP^2 \frac{V_{\phi\phi}}{V}\,, \qquad
\sigma_\phi = \mP^2 \frac{V_{\phi}}{\phi V} \,. 
\eeq
They are given by:
%\be
\begin{align}
\frac{\rd \ln Q}{\rd \Ne } & \simeq \frac{1}{4 -\ceff+ Q(4 + \ceff)} \left(
(4 + 2 \ceff)\epsilon_\phi 
- 2\ceff \eta_\phi 
- 4 (1 -\ceff)\sigma_\phi \right) 
\,, \label{dQ}\\
\frac{\rd \ln T/H}{\rd \Ne } & \simeq \frac{1}{4 -\ceff+ Q(4 + \ceff)} \left(
 \frac{7 - \ceff + Q (5 + \ceff)}{1+Q}\epsilon_\phi 
- 2\eta_\phi 
 -(1-\ceff) \frac{1-Q}{1+Q}\sigma_\phi \right) \,. \label{dTH} \\
\frac{\rd \ln \phi/T}{\rd \Ne } & \simeq \frac{1}{4 -\ceff+ Q(4 + \ceff)} \left(
- \frac{3 + Q}{1+ Q} \epsilon_\phi
+ 2 \eta_\phi 
 -\frac{3+ 5 Q}{1 + Q } \sigma_\phi \right) \,, \label{dphiT}
\end{align}
%\eeq
where: 
\be
\ceff = \frac{3\Upsilon_{\rm LM}}{\Upsilon} + \frac{\Upsilon_{\rm pole}}{\Upsilon} \left(\frac12
  + \frac{\mchi}{T}\right)\,. 
\eeq
Setting $\ceff=3$ in eqs.~\eqref{dQ}-\eqref{dphiT}, we recover the
  evolution equations in the low-momentum-dominated regime given in
  \cite{BasteroGil09}.  

\bibliographystyle{JHEP}
\bibliography{bibliography}

\end{document}